\documentclass[prl, twocolumn, preprintnumbers,amsmath,amssymb,superscriptaddress]{revtex4-1}
\usepackage{graphicx}
\usepackage{dcolumn}
\usepackage{bm}
\usepackage{subfigure}
\usepackage{multirow}
\usepackage{amsmath}
\usepackage{color}
\usepackage{soul}
\usepackage[font=footnotesize, justification=raggedright,singlelinecheck=false]{caption}
\usepackage{mathrsfs}
\usepackage{mathtools}
\usepackage{relsize}

\begin{document}

\title{Non-Equilibrium Structural and Dynamic Behaviors of Polar Active Polymer Controlled by Head Activity}

\author{Jia-Xiang Li}
\affiliation{National Laboratory of Solid State Microstructures and School of Physics, Collaborative Innovation Center of Advanced Microstructures, Nanjing University, Nanjing 210093, P. R. China}

\author{Song Wu}
\affiliation{National Laboratory of Solid State Microstructures and School of Physics, Collaborative Innovation Center of Advanced Microstructures, Nanjing University, Nanjing 210093, P. R. China}

\author{Li-Li Hao}
\affiliation{Research Institute for Biomaterials, Tech Institute for Advanced Materials, College of Materials Science and Engineering, Nanjing Tech University, Nanjing 211816, P. R. China}

\author{Qun-Li Lei}
\email{lql@nju.edu.cn}
\affiliation{National Laboratory of Solid State Microstructures and School of Physics, Collaborative Innovation Center of Advanced Microstructures, Nanjing University, Nanjing 210093, P. R. China}

\author{Yu-Qiang Ma}
\email{myqiang@nju.edu.cn}
\affiliation{National Laboratory of Solid State Microstructures and School of Physics, Collaborative Innovation Center of Advanced Microstructures, Nanjing University, Nanjing 210093, P. R. China}

\begin{abstract}
Thermodynamic behavior of polymer chains out of equilibrium is a fundamental problem in both polymer physics and biological physics. By using molecular dynamics simulation, we discover a non-equilibrium mechanism that controls the conformation and dynamics of polar active polymer, i.e., head activity commands the overall chain activity, resulting in re-entrant swelling of active chains and non-monotonic variation of Flory exponent $\nu$. These intriguing phenomena are results of two  competing non-equilibrium effects arising from the head-controlled railway motion of the chain, i.e., dynamic chain rigidity and the \emph{involution} of chain conformation characterized by the negative bond vector correlation. Moreover, we identify several generic dynamic features of polar active polymers, i.e., \emph{linear} decay of the end-to-end vector correlation function, polymer-size dependent crossover from ballistic to diffusive dynamics, and  diffusion coefficient sensitive to head activity. A simple dynamic theory is proposed to faithfully explain these interesting dynamic phenomena. This sensitive structural and dynamical response of active polymer to its head activity provides us a practical way to control active-agents with applications in biomedical engineering.
\end{abstract}

\maketitle
\section{{\romannumeral1} Introduction}
The scaling theory of polymer introduced by P. G. de Gennes~\cite{ii1979gennes} lays the foundation of polymer physics~\cite{doi1988theory,colby2003polymer}. One of its prediction is the universal scaling behavior of polymer's size, or radius of gyration $R_g$, on monomer number $N$, i.e., $R_g\sim N^\nu$ with $\nu$ the well-known Flory exponent~\cite{Flory1953}. For thermal-equilibrated polymer chain, there are three distinct scaling regimes, i.e., polymer in good solvent ($\nu=0.588$), theta solvent ($\nu=1/2$) and bad solvent ($\nu=1/3$). Nevertheless, how the introduction of non-equilibrium effects modifies these classical predictions is an open question relevant to some key biological processes. For examples, bio-polymer like DNA chromatin in nucleus and actin filaments in cellular cytoskeleton are subjected to propelling forces from either DNA helicases or motor proteins~\cite{Harada1987,Nedelec1997,Julicher2007,Stano2005,Kim2002}, metabolic enzymes can form filamentous membraneless organelles termed cytoophidia that are efficiently transported through 3D complex cellular structures with the help of actin filaments~\cite{Liu2016,Li2018}, polymerlike worms exhibit distinct conformational and dynamical properties with or without driving activity~\cite{Deblais2020,2Deblais2020,Heeremans2022}. These active polymers exhibit complicated self-organized structures and abnormal dynamics that challenge classical polymer physics theory~\cite{Surrey2001,Schaller2011,Vliegenthart2020,Moore2020,Raymond2022,Natali2020,Tanida2020,Cao2022}. On the other side, many artificial active polymers have also been realized, like motility assays of actin filament~\cite{Schaller2010,Sumino2012}, active colloidal polymers~\cite{Dreyfus2005,Hill2014,Biswas2017,Nishiguchi2018}, actuated mechanical chains/ribbons~\cite{Massana-cid2017,Zhang2009,Zhang2009a}, which show potential application value in drug-delivery and soft-body robots. Thus, a deep understanding of the non-equilibrium behaviors of active polymer is of great significance for both biological physics and biomedical engineering~\cite{Winkler2017}.

Active polymers are generally categorized into polar and non-polar. For polar active polymers like actin filaments, active forces are along the backbone of the polymer, thus the total active force has a preferred direction~\cite{Locatelli2021,Mokhtari2019,Prathyusha2018,Isele-Holder2015,Shee2021,Foglino2019,Shi2010}. This is different from the non-polar active polymers like active Brownian chains, where the active forces on  monomers are uncorrelated~\cite{Kaiser2015,Eisenstecken2017,Kaiser2015,Anand2020}. The mobility of the head of polar active polymers can significantly influence the structural and dynamic behaviors of the entire chain. For example, Bourdieu et al. observed in motility assays that the actin filaments rotate like spirals or undulate like flagellums when the head movement is frustrated~\cite{Bourdieu1995}. This phenomenon has inspired many efforts to study active polymers with constrained head~\cite{Camalet1999,DeCanio2017,Ziebert2015,Zheng2023,Wang2022,Martin2019,Chelakkot2014,Chelakkot2021}. Furthermore, head effects also affect the conformation and dynamics of unconstrained active polymers~\cite{Schaller2011a,Hu2022}. Most recently, Patil et al. revealed that the changes in head chirality of the California blackworm are responsible for the rapidly alternation of their self-assembly state~\cite{Patil2023}. Nevertheless, a more general mechanism of head activity-induced conformation and dynamics transition of the polar active polymer remains unknown. Systematic investigations of this issue can contribute to a deeper understanding of the conformation and dynamic behaviors of polar active polymers, including the recently discovered activity-induced coil-to-globule transition~\cite{Bianco2018,Jain2022}.

\begin{figure*}[htbp] 
	\resizebox{140mm}{!}{\includegraphics[trim=0.0in 0.0in 0.0in 0.0in]{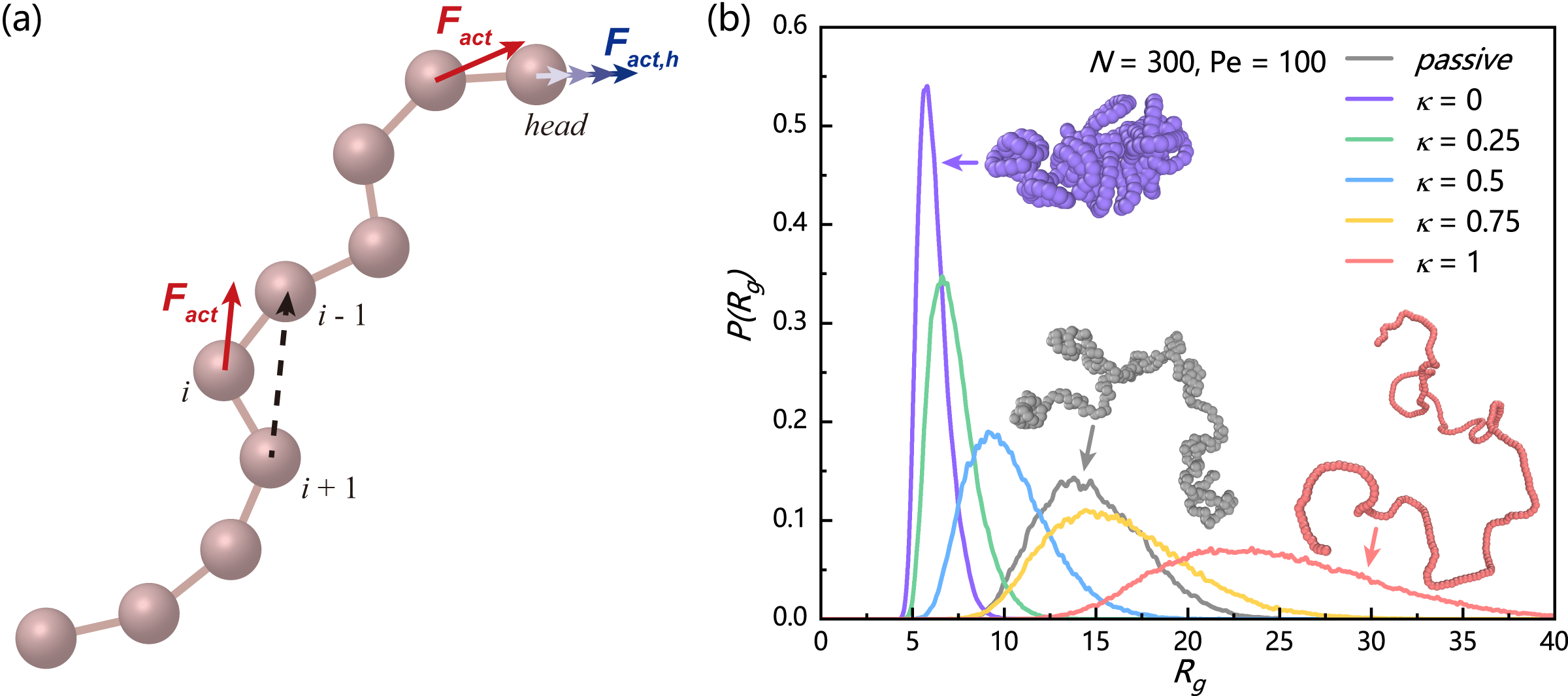} }
	\caption{(a) Schematic representation of the polar active polymer, where red arrows represent active force on backbone monomers and blue arrows with gradient color denote tunable active force on the head monomer. (b) Probability distribution of radius of gyration $P(R_{g})$ for polymers with different head activity strength $\kappa$, where the insets show corresponding chain configurations.}
	\label{fg:scheme}
\end{figure*} 

In this study, we systemically study the structures and dynamics of polar active polymers with controllable head activity. We find that the head activity plays a commanding role for the overall chain activity,  resulting in re-entrant swelling of active polymer and non-monotonic variation of Flory exponent $\nu$, as well as head-activity-controlled polymer dynamics. These intriguing behaviours are results of the competition between two non-equilibrium effects (i.e. dynamic chain rigidity and the \emph{involution} of chain configurations) arising from the head-controlled railway motion of active polymer. Moreover, we also identify several generic dynamic features of polar active polymers, like \emph{linear} rather than exponential decay of the end-to-end vector correlation function,  polymer-size dependent crossover from ballistic to diffusive dynamics, and the head-activity-controlled diffusion coefficient. These dynamic phenomena are explained satisfactorily by  a dynamic theory based on the railway motion.

\begin{figure*}[htbp!] 
	\resizebox{160mm}{!}{\includegraphics[trim=0.0in 0.0in 0.0in 0.0in]{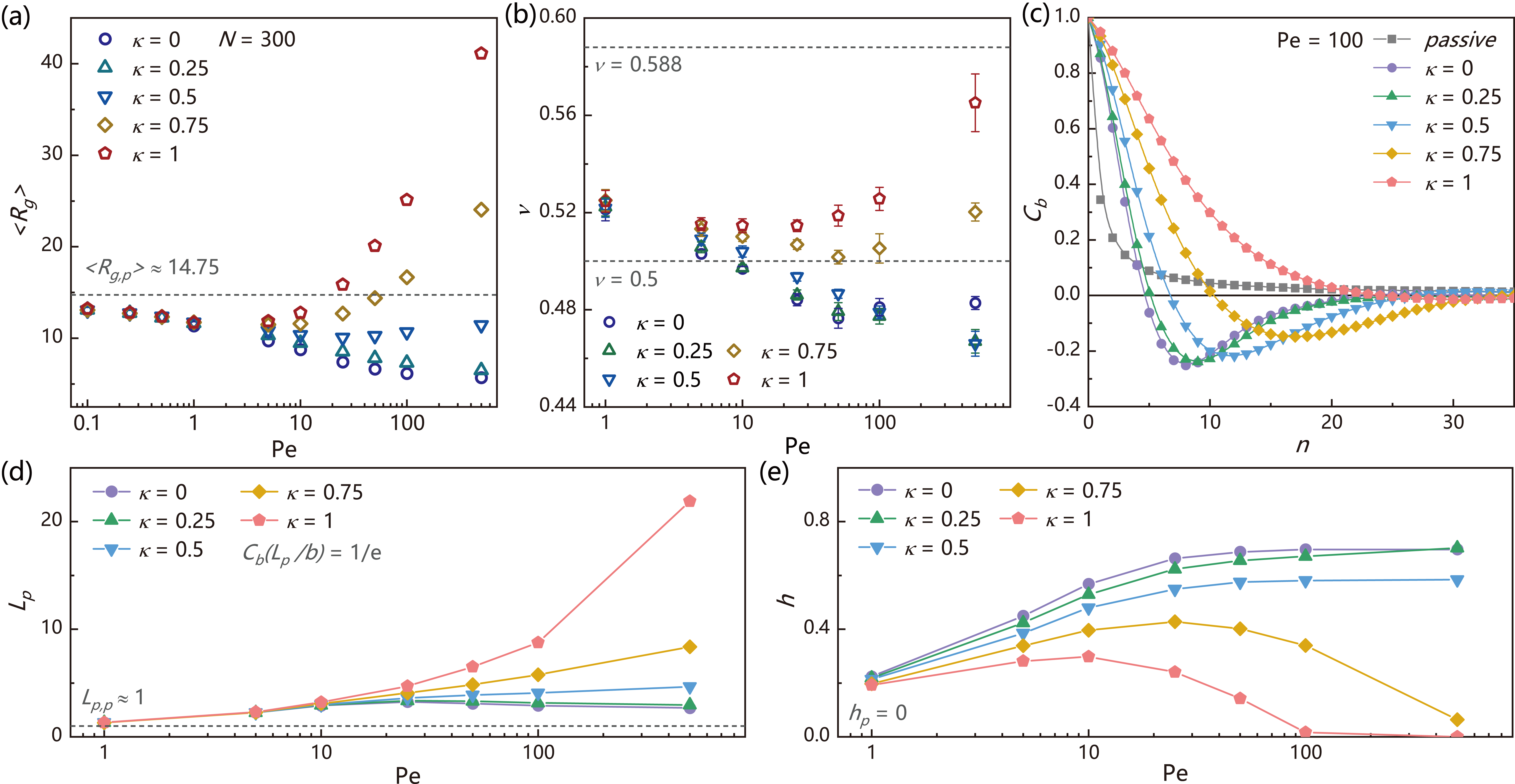} }
	\caption{(a) Average radius of gyration $\langle R_g\rangle$ as a function of activity ${\rm Pe}$ under different head activity strength $\kappa$, where the gray dashed line represents radius of gyration of passive polymers $\langle R_{g,p}\rangle$. (b)  Flory exponent $\nu$ as a function of ${\rm Pe}$. (c) Bond-vector correlation function $C_{b}(n)$ for passive polymer and active polymer under different $\kappa$. (d) Persistence length $L_p$ and (e) negative correlation strength $h$ obtained from $C_{b}(n)$ as functions of ${\rm Pe}$ under different $\kappa$.}
	\label{fg:Rg}
\end{figure*} 

\section{{\romannumeral2} Model and simulations} 
As shown in FIG.~\ref{fg:scheme}(a), we model the active polymer as a bead-spring chain of $N$ monomers with bond length $b$. To ensure conformation stability under self-driving conditions, a modified Kremer-Grest model was used, where the excluded volume interactions are modeled by a WCA-like potential:
\begin{equation}
	\centering
	V_{mm}(r)=\left\{
	\begin{aligned}
		&4\varepsilon\left[\left(\frac{r}{\sigma}\right)^{44}-\left(\frac{r}{\sigma}\right)^{22}\right]+\varepsilon; & r < 2^{1/22}\sigma, \\
		&0; &r\geq2^{1/22}\sigma,
	\end{aligned}
	\right.
	 \label{eq:WCA}
\end{equation}
with $\sigma$ and $\varepsilon$ setting the units of length and energy, and $r$ as the distance between two monomers. Neighboring beads are connected by finitely extensible nonlinear elastic (FENE) potential:
 \begin{equation}
 	\centering
 	V_{b}(r)=\left\{
 	\begin{aligned}
 		&-150\varepsilon\left(\frac{R_{m}}{\sigma}\right)^2{\rm ln}\left[1-\left(\frac{r}{R_{m}}\right)^2\right]; & r < R_{m}, \\
 		&\infty; &r\geq R_{m},
 	\end{aligned}
 	\right.
 	\label{eq:FENE}
\end{equation}
with $R_{m} = 1.05\sigma$. Such a pair of steep potentials severely limits the fluctuations in bond length and can avoid chain crossing in extremely contracted or extended conformations. Based on~\cite{Bianco2018}, we assume that the active force on backbone monomer $i$ has a constant magnitude $F_{act}$ and its direction polarizes to the head monomer ($i=1$) along the local tangent determined by the nearest two monomers, i.e., 
\begin{equation}
	\centering
	\mathbf{F}_{act,i}=F_{act}\mathbf{e}_{i-1,i+1},     ~~~~~~~  i = 2, 3\cdots N-1,
	\label{eq:F_act}
\end{equation}
with $\mathbf{e}_{i,j}=( \mathbf{r}_{i} - \mathbf{r}_{j} )/ \vert \mathbf{r}_{i} - \mathbf{r}_{j} \vert$. The active force on tail monomer  ($i=N$) is set as $\mathbf{F}_{act,N}=F_{act} \mathbf{r}_{N-1, N} / \vert \mathbf{r}_{N-1,N} \vert$, while the active force on head monomer is controlled by the dimensionless head activity strength $\kappa$,
\begin{equation}
	\centering
	\mathbf{F}_{act,1}= \kappa F_{act}    \mathbf{r}_{1, 2} / \vert \mathbf{r}_{1,2} \vert.
	\label{eq:Fact1}
\end{equation}
$\kappa$ can also be understood as the ratio between the head and backbone activities. Standard Langevin dynamics is used to simulate the active polymer in free 3D space:
\begin{equation}
	\centering
	m\frac{d^2\mathbf{r}}{dt^2}=\mathbf{F}_C(\mathbf{r})+\mathbf{F}_R+\mathbf{F}_{act}-\gamma\frac{d\mathbf{r}}{dt} ,
	\label{ea:LGV}
\end{equation}
with $m$ and $\gamma$ setting the mass scale and the friction constant~\cite{Prathyusha2018,Shee2021,Locatelli2021,Isele-Holder2016}. $\mathbf{F}_C$ represents the conservative force (including excluded volume interactions Eq.~(\ref{eq:WCA}) and bonding interactions Eq.~(\ref{eq:FENE})), $\mathbf{F}_R$ denotes the noise with magnitude of $\sqrt{m\gamma k_BT}$, and $\mathbf{F}_{act}$ is the active force applied on the beads, where $k_B$ and $T$ are the Boltzmann constant and temperature, respectively. We have confirmed that Langevin dynamics gives essentially the same results as overdamped Langevin dynamics (Supplemental FIG. S1) but the computational cost is over two orders of magnitude smaller than the latter. In our simulations, $m$ and $\gamma$ are set to unit, hence $k_BT$ controls the thermal energy of the heat bath. The P\'{e}clet number
\begin{equation}
	\centering
	{\rm Pe} \equiv F_{act} b /(k_BT)
	\label{eq:Pe}
\end{equation}
is introduced to quantify the overall activity of the polymer. Without losing generality, we set $F_{act} b = \varepsilon$ as the energy unit of the system, and adjust the P\'{e}clet number by changing thermal energy $k_BT$~\cite{Locatelli2021,Stenhammar2014}. The typical speed of active monomers is $v_0=F_{act} / \gamma$ with $\gamma$ the friction coefficient, and the unit time of the system is chosen as $\tau_0=b/v_0$. Velocity-Verlet algorithm with time step of $\Delta t = 0.001\tau_0$ is used to integrate the equations of polymer motions. At least 160 separate simulations with sufficient time (at least $20N\tau_0$) are performed for each parameter sequences. The control parameters of $N$, ${\rm Pe}$ and $\kappa$ are scanned in the range of $40\leq N\leq 2000$, $0\leq {\rm Pe}\leq 500$ and $0\leq\kappa\leq1$, respectively. All simulations are peformed by a custom modified version of LAMMPS, and all visualizations are produced by OVITO.

\section{{\romannumeral3} Result and Discussion}
\subsection{A. Conformation of polar active polymers} 
The most direct method to investigate the conformation of polymers is the calculation of the gyration radius $R_g$:
\begin{equation}
	\centering
	R_g = \frac{1}{N}\sqrt{\sum_{i=1}^{N}\sum_{j=i}^{N}\langle\left(\mathbf{r}_i-\mathbf{r}_j\right)^2\rangle},
	\label{eq:Rg}
\end{equation}
where $N$ is the number of monomers, $\mathbf{r}_{i.j}$ denotes the position of the monomers. FIG.~\ref{fg:scheme}(b) shows the probability distribution of radius of gyration $P(R_g)$ for active polymers with various head activity under fixed ${\rm Pe}$ and $N$, where $P(R_g)$ for passive polymers is also shown for comparison. In the absence of head activity ($\kappa = 0$), we find strong activity-induced collapse of the polymer chain, consistent with previous studies \cite{Bianco2018,Jain2022}. However, with increasing head activity, active polymer gradually swells and becomes even more expanded than the passive polymer. 

In FIG.~\ref{fg:Rg}(a), we plot the average radius of gyration $\langle R_g\rangle$ as a function of ${\rm Pe}$ under different $\kappa$, where $\langle R_{g,p}\rangle$ for passive polymers is drawn as a dashed line for comparison. For small head activity $ \kappa <0.5$, we find that $\langle R_g\rangle$ shows monotonic decrease with increasing ${\rm Pe}$. Nevertheless, for large head activity, an unusual non-monotonic behaviour of $\langle R_g\rangle$ is observed with a first-stage shrinkage when ${\rm Pe}  \lesssim 10$ and late-state expand at large ${\rm Pe}$. Additionally, the Flory exponent $\nu$ is obtained through plotting $\langle R_g\rangle$ as a function of $N$ (See Supplemental FIG. S2 for more data). In FIG.~\ref{fg:Rg}(b), we show $\nu$ as a function of ${\rm Pe}$ under different $\kappa$ where the reference values of the  random-walk chain ($\nu=0.5$) and self-avoid walks chain ($\nu=0.588$) are drawn as dashed lines. We find that under large head activity, $\nu$ also exhibits a non-monotonic variation with increasing ${\rm Pe}$, which indicates a fundamental change of the fractal dimension of polymer configurations by head activity.

\begin{figure*}[htbp] 
	\resizebox{140mm}{!}{\includegraphics[trim=0.0in 0.0in 0.0in 0.0in]{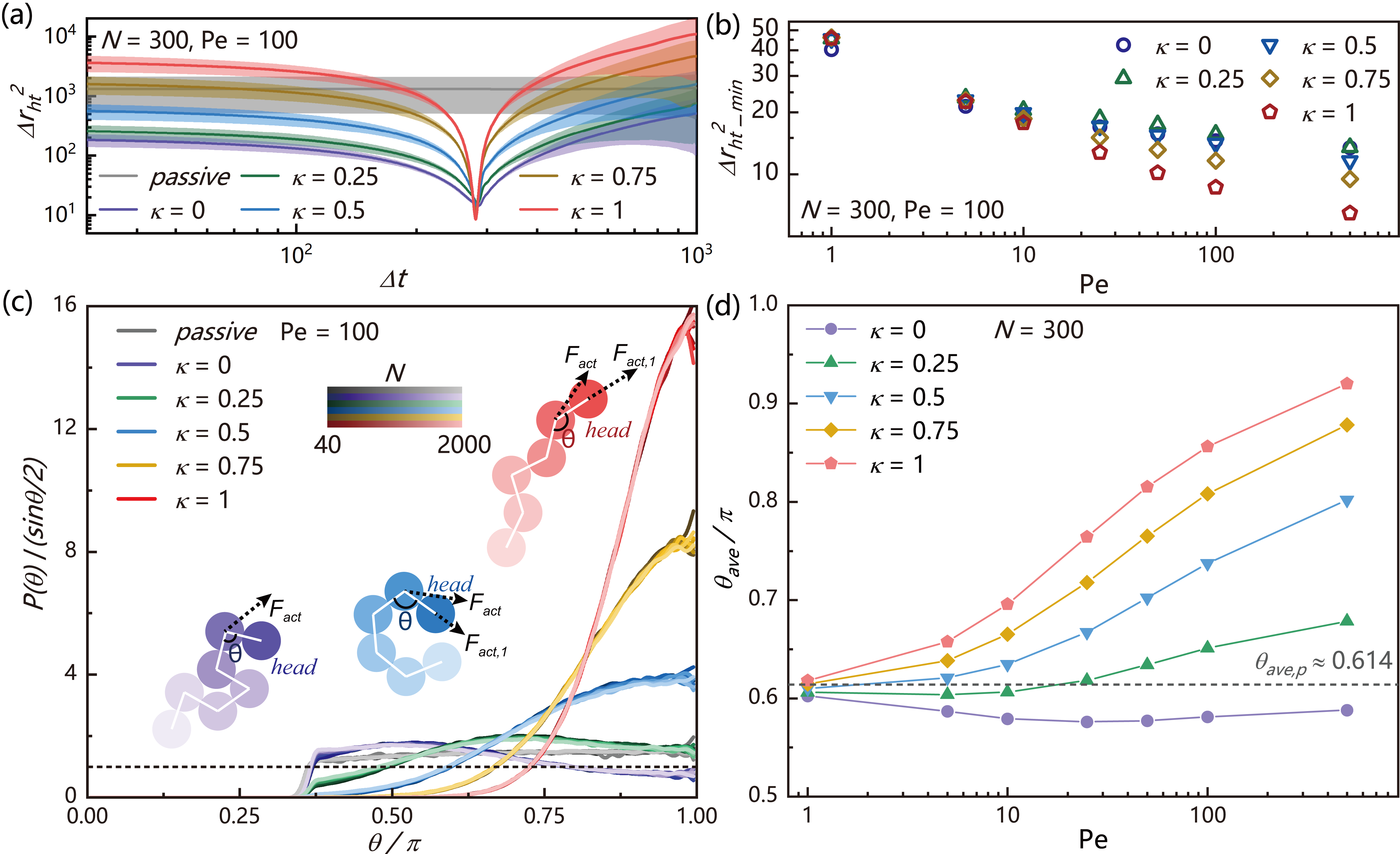} }
	\caption{(a) The correlation function of head-tail position $\Delta r_{ht}^2(\Delta t)$ for polymers with different head activity, where the colored regions represent the error ranges. (b) The minimum value of $\Delta r_{ht}^2$ as a function of ${\rm Pe}$. (c) Probability distribution of the first angle $P(\theta)$ normalized by $sin\theta/2$, where the black dashed line represents $P(\theta)/(sin\theta/2)=1$. Insets: schematic representation of the first angle $\theta$ and a simple mechanical analysis. (d) Averaged $\theta$ as a function of activity ${\rm Pe}$, where the gray dashed line represents the averaged $\theta$ of passive polymers.}
	\label{fg:theta}
\end{figure*} 

To explore these phenomena deeper, we calculate the bond-vector correlation function:
\begin{equation}
	\centering
	C_b(n)=\langle\frac{\mathbf{b}_{i}\cdot \mathbf{b}_{i+n}}{\mathbf{b}_{i}^2}\rangle,
	\label{eq:Cb}
\end{equation}
where $\mathbf{b}_{i}=\mathbf{e}_{i,i+1}$. For typical semi-flexible polymers, $C_b(n)$ decays exponentially. The characteristic correlation length $n^*$ at which $C_b(n^*)$ decays to $e^{-1}$ defines the persistence length $L_p=n^*b$. In FIG.~\ref{fg:Rg}(c), we plot $C_b(n)$ for the passive polymer and active polymers with different $\kappa$ under fixed ${\rm Pe}$. We find that compared with the passive polymer, $C_b(n)$ for active polymers with small $\kappa$ show a longer $L_p$, but with an abnormal negative correlation at intermediate length scale. This negative correlation indicates that the conformation of the active polymer has a strong inward or involution tendency which is associated with the collapsed configuration. Nevertheless, this negative correlation is absent for passive polymers in bad solvent (Supplemental FIG.~S3), which suggests a fundamental difference between these two systems, despite their apparent similarity. As $\kappa$ grows, the persistence length increases significantly, with the negative correlation strength $h$ decrease rapidly to zero. Here, $h$ is defined as the area ratio between negative and positive regions in correlation function. Thus, active polymers with large head activity behave similarly to that of passive semi-flexible chains. This emerging ``dynamic rigidity" is a pure non-equilibrium effect, since the chain still remains mechanically flexible. Unlike semi-flexible passive chains, this dynamic rigidity alone can not predict the conformation of polymers. For example, FIG.~\ref{fg:Rg}(a) shows that even $L_p$ is much larger than that of reference passive polymer, $R_g$ is still smaller than $R_{g,p}$ due to a large negative correlation in $C_b(n)$.

We summarize the behaviors of persistence length $L_p$ and $h$ in FIG.~\ref{fg:Rg}(d) and~\ref{fg:Rg}(e), respectively. One can see that with increasing ${\rm Pe}$, $L_p$ remains around the low value and even decreases for polymer with small $\kappa$, but rises significantly for polymer with large $\kappa$, while $h$ increases monotonically at small $\kappa$ but exhibit strong non-monotonic behavior at large $\kappa$. Therefore, the activity of single head monomer determines the direction of overall active forces acting on the chain: for small head activity, the totally active driving is ``inward", corresponding to a large $h$ (small $L_p$) and a collapsed polymer state, while for large head activity, the totally active forces is ``outward", corresponding to a large $L_p$ (small $h$) and an extended polymer state. 

\begin{figure*}[htbp] 
	\resizebox{140mm}{!}{\includegraphics[trim=0.0in 0.0in 0.0in 0.0in]{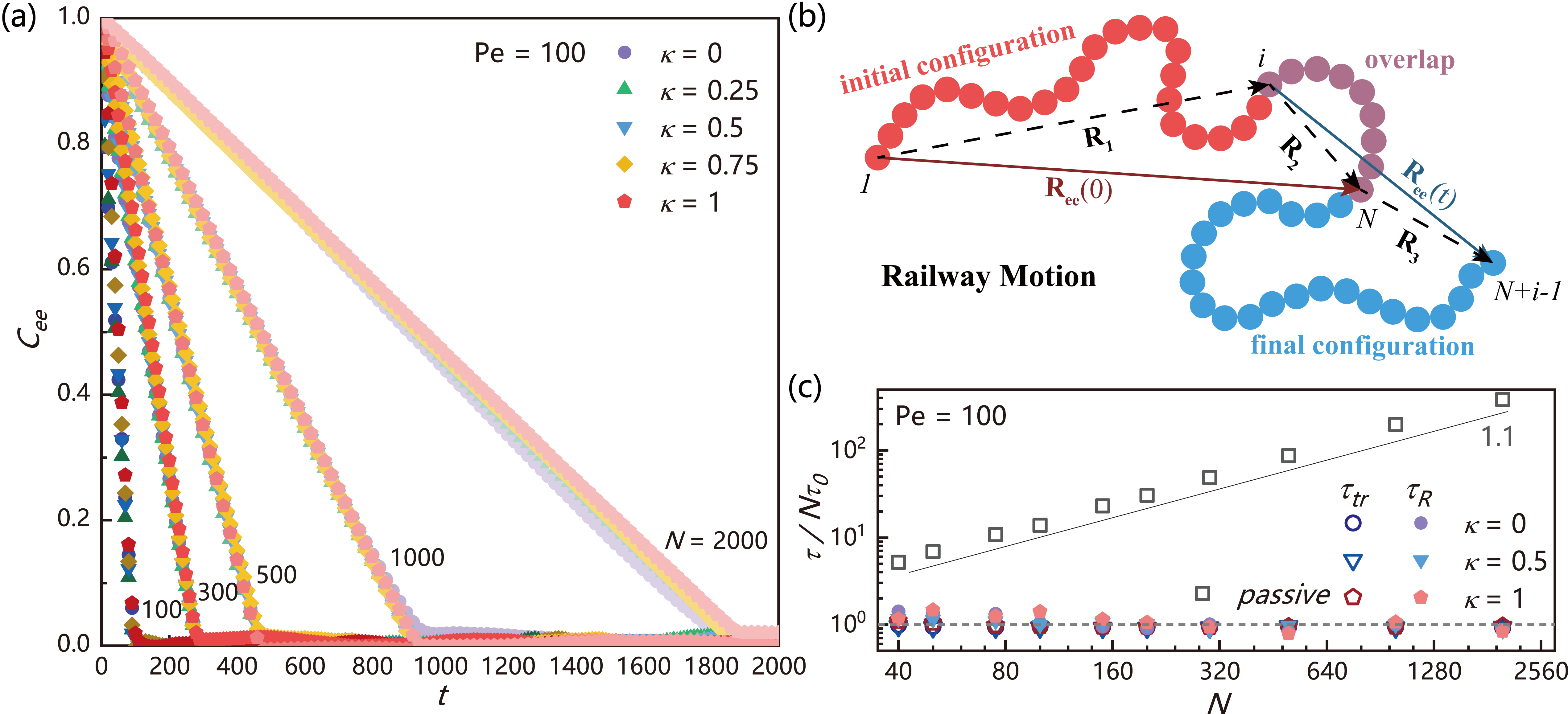} }
 	\caption{(a) Correlation function of end-to-end vector $C_{ee}$ decays linearly with time. (b) Schematic representation of the railway motion. (c) Relaxation time of end-to-end vector correlation function $\tau_{tr}$ and  the critical dynamic transition time $\tau_R$ as functions of chain length. The solid line is guide for eyes and the dashed line represents $\tau= N \tau_0$.}
	\label{fg:Cee}
\end{figure*} 

\subsection{B. Head-controlled railway motion} 
The above abnormal conformation behaviors have a deep connection with the dynamic motion of active polymer. Since the active propelling forces are along the chain's contour, the active polymer adopts \emph{active reptation} motion, which is also referred as ``railway motion"~\cite{Isele-Holder2015} (see Supplemental Movie S1-S5). To quantify such motion, we define a correlation function of head-tail position:
\begin{equation}
	\centering
	\Delta r_{ht}^2(\Delta t) =\langle\left[{\mathbf{r}_1(t)-\mathbf{r}_N(t  + \Delta t )}\right]^2\rangle,
	\label{eq:Rht}
\end{equation}
where the angle bracket represents the time and ensemble average. For perfect railway motion, $\Delta r_{ht}^2$ decays to zero at time $\tau_{ht} \simeq Nb/v_0 =N \tau_0$ and bounces back, which is the time for the tail to reach the original position of the head. For pure diffusive dynamics, $\Delta r_{ht}^2$ will be a flat line determined by polymer's mean squared end-to-end distance $R_e^2\equiv\langle(\mathbf{r}_N-\mathbf{r}_1)^2\rangle$. In  FIG.~\ref{fg:theta}(a), we plot $\Delta r_{ht}^2$ for active polymers with different $\kappa$ under fixed ${\rm Pe}$, where we find  $\Delta r_{ht}^2$ shows pronounce dips at the same $\tau_{ht}$ independent of head activity. We further extract the minimal values of the correlation function of head-tail position, $\Delta r_{ht\_min}^2={\rm min}(\Delta r_{ht}^2(\Delta t))$, to quantify the deviation from the perfect railway motion, and plot them as a function of ${\rm Pe}$ in FIG.~\ref{fg:theta}(b). We find that $\Delta r_{ht\_min}^2$ decreases with  increasing ${\rm Pe}$ and $\kappa$, which indicates that polymers with a higher head activity show relatively stronger railway motion.

For perfect railway motion chain, the direction of the head monomer is the only  degree  of freedom  that determines the chain conformation. This direction can be represented by the bond angle $\theta$ between the first three monomers. In FIG.~\ref{fg:theta}(c), we give the angle distribution $P(  \theta  )$ for active polymers with different $\kappa$ along with the  passive polymer under varying polymer lengths. We find that for the active polymer without head activity, the peak of $P(  \theta  )$ locates at a small value around $0.5\pi$, indicating the averaged bent configuration of polymer head. With increasing $\kappa$, the polymer head takes a more straight configuration. These results are independent of polymer length, indicating a similar way of the head activity affecting the polymer contour. Simple mechanical analysis in the inset of FIG.~\ref{fg:theta}(c) shows that when the propulsion force of the head monomer is smaller than the second monomer, the pushing force from back will induce a ``torque" that makes the bond $\mathbf r_{1,2}$ bending inward. One the contrary, when the propulsion force of the head monomer is large enough, the head monomer will pull the three monomers straight. For railway motion chain, $P(  \theta)$ determines the bond angle distribution of the whole chain. Thus, an increasing average $\theta$ corresponds to an increasing persistence length $L_p$ and a decreasing involution degree $h$ of polymer. In FIG.~\ref{fg:theta}(d), we summarized the average value of $\theta$, which is qualitatively agreed with the behaviour of $L_p$ and $h$ in  FIG.~\ref{fg:Rg}(d) and \ref{fg:Rg}(e).

\begin{figure*}[htbp] 
	\resizebox{140mm}{!}{\includegraphics[trim=0.0in 0.0in 0.0in 0.0in]{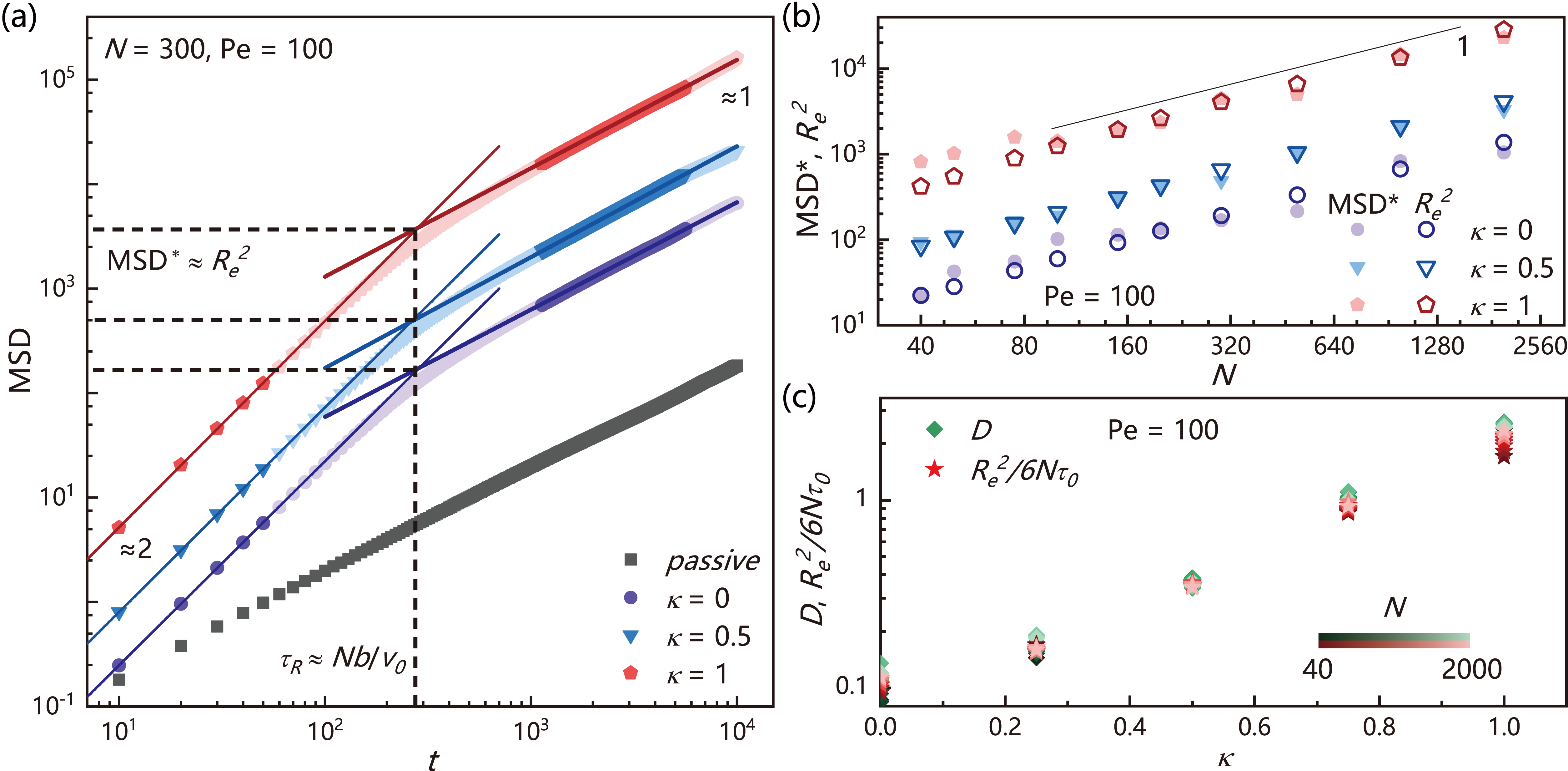} }
	\caption{(a) $\rm MSD(t)$ of the polymers, where solid lines are linear fitting at different dynamic regimes and dashed lines indicate the critical $\rm MSD^*$ and $\tau_R$. (b) $\rm MSD^*$ and $R_e^2$ as functions of chain length $N$. (c) Diffusion coefficient $D$ and $R_e^2/6N\tau_{0}$ as functions of $\kappa$ under different polymer lengths.}
	\label{fg:D}
\end{figure*} 

\subsection{C. Dynamics of polar active polymers} 
To further explore the impact of head activity on the dynamics of polar active polymer, we calculate the end-to-end vector correlation function
\begin{equation}
		\centering
		C_{ee}(t)=\frac{\mathbf{R_e}(0)\cdot\mathbf{R_e}(t)}{\mathbf{R_e}(0)^2},
		\label{eq:Cee1}
\end{equation}
and compare $C_{ee}(t)$ for polymers with different head activity $\kappa$ and polymer length $N$ in FIG.~\ref{fg:Cee}(a). A surprising finding is that all $C_{ee}(t)$ exhibit \emph{linear} decay rather than exponential decay as for the passive polymer (Supplemental FIG.~S4)~\cite{Azuma1999,Descas2004}.  Moreover, we find $C_{ee}(t)$ is insensitive to the head activity $\kappa$ and overall activity ${\rm Pe}$ but only depends on the chain length $N$. In FIG.~\ref{fg:Cee}(c), we plot the characteristic decay time $\tau_{tr}$ of $C_{ee}(t)$ (for active polymers $C_{ee}(\tau_{tr})=0$ while for passive polymers $C_{ee}(\tau_{tr})=1/e$) as a function of $N$, we find all data from active polymer falling around the line of $\tau= N \tau_0$. This behavior is also distinct from the passive polymer for which $\tau_{tr} \propto N^2$.
 
Such abnormal dynamics can be explained theoretically based on railway motion of the polymer chain. As shown in FIG.~\ref{fg:Cee}(b), assuming the initial and final configurations of the polymer on the same ``railway" have an overlap part with end-to-end vector $\mathbf R_2$, thus we have ${\mathbf R_e}(0)= \mathbf R_1 + \mathbf R_2 $ and ${\mathbf R_e}(t)= \mathbf R_2+ \mathbf R_3 $, where $\mathbf R_1$ and $\mathbf R_3$ are end-to-end vectors of the two non-overlapped parts. Since $\mathbf R_1$, $\mathbf R_2$, $\mathbf R_3$ are statistically independent, we have $\left\langle\mathbf{R}_i\cdot\mathbf{R}_j\right\rangle=0$ for $i \neq j$. Combining the nearly linear dependence of $\mathbf{R}_e^2$ on $N$ shown in FIG.~\ref{fg:D}(b), one can get
\begin{equation}
	\centering
	C_{ee}(t)\approx\frac{\mathbf{R}_2^2}{\mathbf{R_e}(0)^2}\approx\frac{Nb-v_0t}{Nb}=1-\frac{t}{N \tau_0},\quad t< N \tau_0 ,
	\label{eq:Cee2}
\end{equation}
which suggests $\tau_{tr}=N \tau_0$. Note that this result is independent of the shape of ``railway" controlled by head activity. 

The railway motion also results in interesting polymer diffusion dynamics. Starting from the railway motion hypothesis in FIG.~\ref{fg:Cee}(b), one can get the expression of the center-of-mass of the polymer as
\begin{equation}
	\centering
	\mathbf{r}_{cm}(0)=\frac{1}{N}\sum_{i=1}^{N}\mathbf{r}_i,\quad \mathbf{r}_{cm}(t)=\frac{1}{N}\sum_{i=t}^{N+t}\mathbf{r}_i.
	\label{eq:rcm}
\end{equation}
When $t<N\tau_0$, the displacement of the center-of-mass could be rewritten as:
\begin{equation}
	\centering
	\Delta\mathbf{r}_{cm}(t)=\frac{1}{N}\left(\sum_{i=t}^{N+t}\mathbf{r}_i-\sum_{i=1}^{N}\mathbf{r}_i\right) =\frac{1}{N}\left(\mathbf{r}_{N+i}-\mathbf{r}_i\right).
	\label{eq:drcm}
\end{equation}
Using an approximation of $\mathbf{r}_{N+i} - \mathbf{r}_i \approx \mathbf{R_e}(i)$ and considering continuous motion, we get
\begin{equation}
	\centering
	\Delta\mathbf{r}_{cm}(t)=\frac{1}{N}\sum_{i=1}^t\mathbf{R_e}(i)=\frac{1}{N\tau_0}\int_0^t\mathbf{R_e}(t){\rm d}t.
	\label{eq:com}
\end{equation}
Thus, the mean square displacement (MSD) of polymer center-of-mass has the formation of
\begin{align}
	{\rm MSD} (t) &\equiv 	\Delta\mathbf{r}_{cm}^2(t) \nonumber \\
		&=\frac{1}{N^2\tau_0^2}\int_0^t\int_0^t\mathbf{R_e}(t_1)\mathbf{R_e}(t_2){\rm d}t_1{\rm d}t_2
	\label{eq:MSD1}
\end{align}
when $\left|t_2-t_1\right|\leq N\tau_0$. Combining the linear decay of $C_{ee}$ (Eq.~(\ref{eq:Cee2})), we get
\begin{equation}
	\centering
	{\rm MSD}(t) = \frac{1}{N^2\tau_0^2}\int_0^t\int_0^t\mathbf{R}_{\mathbf{e}}^2(t_1)\left(1-\frac{\left|t_2-t_1\right|}{N\tau_0}\right){\rm d}t_1{\rm d}t_2.
	\label{eq:MSD2}
\end{equation}
Considering the long-time limit $t\gg N\tau_0$, the internal integral can be completely computed and has symmetric properties. Hence, 
\begin{align}
	{\rm MSD}(t) 
	&=\frac{2}{N^2\tau_0^2}\int_0^t\int_{t_1}^{t_1+N\tau_0}\mathbf{R}_{\mathbf{e}}^2(t_1)\left(1-\frac{t_2-t_1}{N\tau_0}\right){\rm d}t_1{\rm d}t_2  \nonumber \\
	&=\frac{2}{N^2\tau_0^2}\int_0^t\int_0^{N\tau_0}\mathbf{R}_{\mathbf{e}}^2(t_1)\left(1-\frac{t_3}{N\tau_0}\right){\rm d}t_1{\rm d}t_3 \nonumber \\
	&=\frac{2}{N^2\tau_0^2}\int_0^t\mathbf{R}_{\mathbf{e}}^2(t_1)\frac{N\tau_0}{2}{\rm d}t_1,
	\label{eq:MSD3}
\end{align}
where $t_3 = t_2 - t_1$. During the whole time $\mathbf{R}_{\mathbf{e}}^2(t_1)$ is randomly fluctuated around the average end-to-end distance $R_e^2$, hence the integral could be directly calculated as
\begin{equation}
	{\rm MSD}(t)=\frac{R_e^2}{N\tau_0}t, \quad t\gg N\tau_0.
	\label{eq:MSD4}
\end{equation}
So now we have a strict expression for ${\rm MSD}$ on large time-scale. When the situation is opposite, $t\ll N\tau_0$, the internal integral of Eq.~(\ref{eq:MSD2}) cannot be completely computed. Considering the symmetry of $t_1>t_2$ and $t_1<t_2$, we have
\begin{align}
	{\rm MSD}(t) &=\frac{2}{N^2\tau_0^2}\int_0^t\int_0^{t_1}\mathbf{R}_{\mathbf{e}}^2(t_1)\left(1-\frac{t_1-t_2}{N\tau_0}\right){\rm d}t_1{\rm d}t_2 \nonumber \\
	&=\frac{2}{N^2\tau_0^2}\int_0^t\mathbf{R}_{\mathbf{e}}^2(t_1)\left(t_1-\frac{t_1^2}{2N\tau_0}\right){\rm d}t_1.
	\label{eq:MSD5}
\end{align}
When $t\ll N\tau_0$, $\mathbf{R}_{\mathbf{e}}^2(t_1)$ has little fluctuation and nearly maintains the initial value with averages of  $R_e^2$, hence MSD has an approximate value of
\begin{equation}
	{\rm MSD}(t)\approx\frac{2R_e^2}{N^2\tau_0^2}\left(\frac{t^2}{2}-\frac{t^3}{6N\tau_0}\right), \quad t\ll N\tau_0.
	\label{eq:MSD6}
\end{equation}
In summary, the expression of the mean squared displacement of center-of-mass of an active polymer has the formation of
\begin{equation}
	{\rm MSD} (t) \simeq \left\{
		\begin{aligned}
			&\frac{R_e^2}{N^2 \tau_0^2 } {t^2},   ~~~~~ \quad & t \ll N \tau_0, \\
			&\frac{R_e^2 }{N \tau_0}t,      ~~~~~    \quad & t  \gg N \tau_0.
		\end{aligned}
	\right.
	\label{eq:MSD}
\end{equation}
Eq.~(\ref{eq:MSD}) predicts a ballistic dynamics regime (${\rm MSD}\sim t^2$) and a diffusion dynamics regime (${\rm MSD}\sim t^1$) separated by the characteristic time $\tau_R = N\tau_0$ and characteristic ${\rm MSD}^*(\tau_R)\simeq {R_e^2}$. In FIG.~\ref{fg:D}(a) and Supplemental FIG.~S5, we plot the $\rm MSD(t)$ for active polymers with different $\kappa$, where the reference $\rm MSD(t)$ of the passive polymer is shown in gray color. We find that the simulation results agree well with our theoretical prediction. Especially $\rm MSD^*(\tau_R)$ match with $R_e^2$ excellently when varying the head activity and polymer length (see FIG.~\ref{fg:D}(b)). Furthermore, based on Stokes-Einstein relation ${\rm MSD}=6Dt$ and Eq.~(\ref{eq:MSD}), we can obtain the diffusion coefficient
\begin{equation}
	D=\frac{R_e^2}{6N\tau_0}.
	\label{eq:D}
\end{equation}
Since $R_e^2 \sim N $, the diffusion coefficient is independent of chain length, which is in contrast with the behavior of classical Rouse chain $D \sim \frac{1}{N}$~\cite{Tejedor2019} and consistent with previous work~\cite{Bianco2018}. These predictions are also confirmed by the simulation results shown in FIG.~\ref{fg:D}(c) and Supplemental FIG.~S6, at which $D$ is independent of $N$ and matches with $R_e^2/6N\tau_0$ very well under different $\kappa$. Moreover, from FIG.~\ref{fg:D}(c), one can see that the conformation and dynamics are coupled: the diffusion coefficient of an extended active polymer with large head activity can be an order of magnitude larger than the collapsed active polymer with small head activity. Finally, we emphasize another scaling relation hidden in Eq.~(\ref{eq:D}), $D\sim\tau_0^{-1}$. Since $\tau_0$ is defined as $\tau_0=b/v_0$ with $v_0=F_{act}/\gamma$, we have $D\sim F_{act}$, which provids a connection between dynamics and activity strength. This correlation has been a subject of continuous debate and contention in literatures~\cite{Anand2018,Bianco2018,Mokhtari2019}. The dynamic theory developed here is based on railway motion, hence it is valid for all polar active polymers exhibiting railway motion, regardless of the model (Supplemental FIG.~S7). This provides us an efficient way to control the conformation and dynamics of active polymer, which may have application values in drug delivery and biomedical engineering~\cite{Sasaki2014,Vuijk2021,Isele-Holder2016}. Recently, Philipps et al. submitted an analytical study on polar active polymers, which reports dynamic equations similar to Eq.~(\ref{eq:D})~\cite{Philipps2022}.
 
\section{{\romannumeral3} Conclusion}
In conclusion, we find a general non-equilibrium mechanism that head activity commands the overall active forces on the polar active polymer performing railway motion, which leads to unusual conformation and dynamics behaviors of the polymer chain. We demonstrate that the low head activity favors a bent head configuration, which causes a strong involution  and the collapsing of polymer chain, while the high head activity helps the polymer straighten its head, resulting in an emerging ``dynamic rigidity" and more extended chain conformation. The competition between the two non-trivial non-equilibrium effects leads to the re-entrant swelling of polar active polymers and non-monotonic variation of the Flory exponent $\nu$. We also find many interesting dynamic features for polar active polymers, like a linear-decay of end-to-end vector correlation function $C_{ee}$, a polymer-size dependent crossover from ballistics to diffusion dynamics, as well as a length-independent diffusion coefficient controlled by the head-activity. All these features are explained well by our dynamic theory. Our findings are not only meaningful for understanding the conformational and dynamical behaviors of complicated active polymer aggregations, but also suggestive for designing efficient polymer-based drug delivery systems. In this work, the hydrodynamic interactions (HI) are not considered because the unique railway motion pattern of polar active polymers allows them to exhibit a qualitatively similar behavior regardless of the presence or absence of HI~\cite{Jiang2014,Jiang2014a}. Notwithstanding, the effects of HI on polar active polymers remain a highly significant research topic that requires further investigations.

\section{Acknowledgments}  We thank Prof. Paolo Malgaretti in Max Planck Institute for helpful discussions. This work is supported by the National Natural Science Foundation of China (No.12104219).

\bibliographystyle{apsrev4-2}
\bibliography{main}

\end{document}